\documentclass[journal]{IEEEtran}
\usepackage{url}
\usepackage[utf8]{inputenc}
\usepackage{xcolor}
\usepackage{amsmath}
\usepackage{amssymb}

\usepackage[acronyms,nonumberlist,nopostdot,nomain,nogroupskip]{glossaries}
\usepackage{tablefootnote}
\usepackage{booktabs}
\usepackage{tabularx}
\usepackage{tikz}
\usepackage{pgfplots}
\pgfplotsset{compat=newest} 
\pgfplotsset{plot coordinates/math parser=false} 
\newlength\fheight
\newlength\fwidth
\usetikzlibrary{plotmarks,patterns,decorations.pathreplacing,backgrounds,calc,arrows,arrows.meta,spy,matrix}
\usepgfplotslibrary{patchplots,groupplots}
\usepackage{tikzscale}
\usepackage{hyperref}
\usepackage{siunitx}

\usepackage{multirow}
\usepackage{tkz-kiviat}

\usepackage[font=scriptsize]{subcaption}
\usepackage[font=footnotesize]{caption}

\usepackage{mathtools}

\usepackage{dblfloatfix}    
\usepackage{colortbl}

\usepackage{makecell}
\usepackage{diagbox}
\usepackage{tikz-qtree}
\usetikzlibrary{trees} 
\usepackage{cite}
\usepackage{bbold}
\usepackage{bbm}
\usepackage[outdir=./]{epstopdf}

\usepackage{array}
\newcolumntype{?}{!{\vrule width 1.5pt}}
\newcolumntype{P}[1]{>{\centering\arraybackslash}p{#1}}
\usepackage{makecell}
\usepackage{mdframed}
\usepackage[many]{tcolorbox}
\usepackage{enumitem}
\usepackage{booktabs}

\newtcbox{\mybox}[1][]{nobeforeafter,math upper,tcbox raise base,
  enhanced,frame hidden,boxrule=0pt,interior style={top color=green!10!white,
  bottom color=green!10!white,middle color=green!50!yellow},
  fuzzy halo=1pt with green,drop large lifted shadow,#1}

\usepackage{siunitx}
\usetikzlibrary{fadings}

\tikzfading[name=middle,
            top color=transparent!100,
            bottom color=transparent!100,
            middle color=transparent!20]

\usetikzlibrary{arrows,automata,calc,shapes, positioning,shadows,shadows.blur,shapes.geometric}

\newacronym{3gpp}{3GPP}{3rd Generation Partnership Project}
\newacronym{adc}{ADC}{Analog to Digital Converter}
\newacronym{5g}{5G}{5th generation}
\newacronym{6g}{6G}{6th generation}
\newacronym{aimd}{AIMD}{Additive Increase Multiplicative Decrease}
\newacronym{am}{AM}{Acknowledged Mode}
\newacronym{amc}{AMC}{Adaptive Modulation and Coding}
\newacronym{aqm}{AQM}{Active Queue Management}
\newacronym{awgn}{AGWN}{Additive White Gaussian Noise}
\newacronym{balia}{BALIA}{Balanced Link Adaptation}
\newacronym{bdp}{BDP}{Bandwidth-Delay Product}
\newacronym{bf}{BF}{beamforming}
\newacronym{cc}{CC}{Congestion Control}
\newacronym{cdf}{CDF}{Cumulative Distribution Function}
\newacronym{cn}{CN}{Core Network}
\newacronym{cr}{CR}{cognitive radio}
\newacronym{cqi}{CQI}{Channel Quality Information}
\newacronym{cp}{CP}{Control Plane}
\newacronym{csirs}{CSI-RS}{Channel State Information - Reference Signal}
\newacronym{dc}{DC}{Dual Connectivity}
\newacronym{rb}{RB}{Resource Block}
\newacronym{dce}{DCE}{Direct Code Execution}
\newacronym{dci}{DCI}{Downlink Control Information}
\newacronym{udp}{UDP}{User Datagram Protocol}
\newacronym{dl}{DL}{Downlink}
\newacronym{dmr}{DMR}{Deadline Miss Ratio}
\newacronym{dmrs}{DMRS}{DeModulation Reference Signal}
\newacronym{e2e}{E2E}{End-to-End}
\newacronym{ppp}{PPP}{Poission Point Process}
\newacronym{si}{SI}{Study Item}
\newacronym{ecn}{ECN}{Explicit Congestion Notification}
\newacronym{edf}{EDF}{Earliest Deadline First}
\newacronym{enb}{eNB}{eNodeB}
\newacronym{epc}{EPC}{Evolved Packet Core}
\newacronym{es}{ES}{Edge Server}
\newacronym{cav}{CAV}{Connected and Autonomous Vehicle}
\newacronym{fdma}{FDMA}{Frequency Division Multiple Access}
\newacronym{fdd}{FDD}{Frequency Division Duplexing}
\newacronym{upa}{UPA}{Uniform Planar Array}
\newacronym[firstplural=Radio Access Technologies (RATs)]{rat}{RAT}{radio access technology}
\newacronym[firstplural=Radio Access Technology (RTs)]{rt}{RT}{Radio Technology}
\newacronym{fs}{FS}{Fast Switching}
\newacronym{isd}{ISD}{inter-site distance}
\newacronym{ftp}{FTP}{File Transfer Protocol}
\newacronym{gnb}{gNB}{Next Generation Node Base}
\newacronym{harq}{HARQ}{Hybrid Automatic Repeat reQuest}
\newacronym{hetnet}{HetNet}{Heterogeneous Network}
\newacronym{hh}{HH}{Hard Handover}
\newacronym{hol}{HOL}{Head-of-Line}
\newacronym{ia}{IA}{Initial Access}
\newacronym{imt}{IMT}{International Mobile Telecommunication}
\newacronym{iot}{IoT}{Internet of Things}
\newacronym{los}{LOS}{line of sight}
\newacronym{lte}{LTE}{Long Term Evolution}
\newacronym{m2m}{M2M}{Machine to Machine}
\newacronym{mac}{MAC}{Medium Access Control}
\newacronym{mc}{MC}{Multi-Connectivity}
\newacronym{mcs}{MCS}{modulation and coding scheme}
\newacronym{mec}{MEC}{Mobile Edge Cloud}
\newacronym{mi}{MI}{Mutual Information}
\newacronym{mimo}{MIMO}{Multiple Input Multiple Output}
\newacronym{mmwave}{mmWave}{millimeter wave}
\newacronym{mptcp}{MPTCP}{Multipath TCP}
\newacronym{mr}{MR}{Maximum Rate}
\newacronym{mss}{MSS}{Maximum Segment Size}
\newacronym{mtd}{MTD}{Machine-Type Device}
\newacronym{mtu}{MTU}{Maximum Transmission Unit}
\newacronym{nfv}{NFV}{Network Function Virtualization}
\newacronym{vnf}{VNF}{ Virtualization Network Function}
\newacronym{sdn}{SDN}{software defined networking}
\newacronym{nlos}{NLOS}{Non Line of Sight}
\newacronym{nlosb}{NLOSb}{Building Non Line of Sight}
\newacronym{nlosv}{NLOSv}{Vehicle Non Line of Sight}
\newacronym{nr}{NR}{New Radio}
\newacronym{ofdm}{OFDM}{Orthogonal Frequency Division Multiplexing}
\newacronym{pdcch}{PDCCH}{Physical Downlonk Control Channel}
\newacronym{pdcp}{PDCP}{Packet Data Convergence Protocol}
\newacronym{pdsch}{PDSCH}{Physical Downlink Shared Channel}
\newacronym{pdu}{PDU}{Packet Data Unit}
\newacronym{pf}{PF}{Proportional Fair}
\newacronym{pgw}{PGW}{Packet Gateway}
\newacronym{phy}{PHY}{Physical}
\newacronym{pbch}{PBCH}{Physical Broadcast Channel}
\newacronym[plural=\gls{mme}s,firstplural=Mobility Management Entities (MMEs)]{mme}{MME}{Mobility Management Entity}
\newacronym{prb}{PRB}{Physical Resource Block}
\newacronym{pss}{PSS}{Primary Synchronization Signal}
\newacronym{pucch}{PUCCH}{Physical Uplink Control Channel}
\newacronym{pusch}{PUSCH}{Physical Uplink Shared Channel}
\newacronym{rach}{RACH}{Random Access Channel}
\newacronym{ran}{RAN}{radio access network}
\newacronym{red}{RED}{Random Early Detection}
\newacronym{rf}{RF}{radio frequency}
\newacronym{rlc}{RLC}{Radio Link Control}
\newacronym{rlf}{RLF}{Radio Link Failure}
\newacronym{rrc}{RRC}{Radio Resource Control}
\newacronym{rrm}{RRM}{Radio Resource Management}
\newacronym{rr}{RR}{Round Robin}
\newacronym{rs}{RS}{Remote Server}
\newacronym{rsrp}{RSRP}{Reference Signal Received Power}
\newacronym{rss}{RSS}{Received Signal Strength}
\newacronym{rtt}{RTT}{Round Trip Time}
\newacronym{rw}{RW}{Receive Window}
\newacronym{rx}{RX}{Receiver}
\newacronym{sa}{SA}{standalone}
\newacronym{sack}{SACK}{Selective Acknowledgment}
\newacronym{sap}{SAP}{Service Access Point}
\newacronym{sch}{SCH}{Secondary Cell Handover}
\newacronym{scoot}{SCOOT}{Split Cycle Offset Optimization Technique}
\newacronym{sdma}{SDMA}{Spatial Division Multiple Access}
\newacronym{sinr}{SINR}{Signal to Interference plus Noise Ratio}
\newacronym{sm}{SM}{Saturation Mode}
\newacronym{snr}{SNR}{Signal to Noise Ratio}
\newacronym{son}{SON}{Self-Organizing Network}
\newacronym{ss}{SS}{Synchronization Signal}
\newacronym{srs}{SRS}{Sounding Reference Signal}
\newacronym{sss}{SSS}{Secondary Synchronization Signal}
\newacronym{tb}{TB}{Transport Block}
\newacronym{tcp}{TCP}{Transmission Control Protocol}
\newacronym{tdd}{TDD}{Time Division Duplexing}
\newacronym{tdma}{TDMA}{Time Division Multiple Access}
\newacronym{tfl}{TfL}{Transport for London}
\newacronym{tm}{TM}{Transparent Mode}
\newacronym{prr}{PRR}{Packet Reception Ratio}
\newacronym{trp}{TRP}{Transmitter Receiver Pair}
\newacronym{tti}{TTI}{Transmission Time Interval}
\newacronym{ttt}{TTT}{Time-to-Trigger}
\newacronym{tx}{TX}{Transmitter}
\newacronym{ue}{UE}{User Equipment}
\newacronym{ul}{UL}{Uplink}
\newacronym{uml}{UML}{Unified Modeling Language}
\newacronym{um}{UM}{Unacknowledged Mode}
\newacronym{utc}{UTC}{Urban Traffic Control}
\newacronym{vm}{VM}{Virtual Machine}
\newacronym{rsrq}{RSRQ}{Reference Signal Received Quality}
\newacronym{rssi}{RSSI}{Received Signal Strength Indicator}
\newacronym{crs}{CRS}{Cell Reference Signal}
\newacronym{v2v}{V2V}{Vehicle-to-Vehicle}
\newacronym{v2i}{V2I}{Vehicle-to-Infrastructure}
\newacronym{v2n}{V2N}{Vehicle-to-Network}
\newacronym{v2x}{V2X}{Vehicle-to-Everything}
\newacronym{vn}{VN}{Vehicular Node}
\newacronym{dsrc}{DSRC}{Dedicated Short Range Communication}
\newacronym{ci}{CI}{context information}
\newacronym{voi}{VoI}{value of information}
\newacronym{gps}{GPS}{Global Positioning System}
\newacronym{qos}{QoS}{Quality of Service}
\newacronym{qoe}{QoE}{Quality of Experience}
\newacronym{ml}{ML}{machine learning}
\newacronym{ahp}{AHP}{Analytic Hierarchy Process}
\newacronym{lidar}{LIDAR}{Light Detection and Ranging}
\newacronym{sumo}{SUMO}{Simulation of Urban MObility}
\newacronym{wave}{WAVE}{Wireless Access in Vehicular Environment}
\newacronym{c-its}{C-ITS}{Connected Intelligent Transportation System}
\newacronym{dash}{DASH}{Dynamic Adaptive Streaming over HTTP}
\newacronym{http}{HTTP}{HyperText Transfer Protocol}
\newacronym{nt}{NT}{non-terrestrial}
\newacronym{ntc}{NTC}{non-terrestrial communication}
\newacronym{ntn}{NTN}{non-terrestrial network}
\newacronym{haps}{HAPS}{high altitude platform station}
\newacronym{leo}{LEO}{Low Earth Orbit}
\newacronym{meo}{MEO}{Medium Earth Orbit}
\newacronym{geo}{GEO}{Geostationary Earth Orbit}
\newacronym{uav}{UAV}{unmanned aerial vehicle}
\newacronym{nsat}{nSAT}{Nanosatellite}
\newacronym{ehf}{EHF}{extremely high-frequency}
\newacronym{ioe}{IoE}{Internet of Everyone}
\newacronym{gan}{GaN}{Gallium Nitride}
 \newacronym{kpi}{KPI}{key performance indicator}
 \newacronym{mno}{MNO}{mobile network operator}
 \newacronym{plc}{PLC}{power line communication}
 \newacronym{isp}{ISP}{Internet service provider}
 \newacronym{sdr}{SDR}{software defined radio}
 \newacronym{iab}{IAB}{integrated access and backhaul}
 \newacronym{fso}{FSO}{free space optical}
 \newacronym{eon}{EON}{elastic optical network}
 \newacronym{ai}{AI}{artificial intelligence}
 \newacronym{noma}{NOMA}{non-orthogonal multiple access}
 \newacronym{irs}{IRS}{intelligent reflecting surface}
 \newacronym{oam}{OAM}{Operations, administration and management}
\newacronym{ws}{WS}{white space}
\newacronym{d2d}{D2D}{device-to-device}
\newacronym{owc}{OWC}{optical wireless communication}
\newacronym{vlc}{VLC}{visible light communication}
\newacronym{vr}{VR}{virtual reality} \newacronym{ar}{AR}{augmented reality}
\newacronym{3d}{3D}{three-dimensional}
\newacronym{lbt}{LBT}{listen-before-talk}
\newacronym{eh}{EH}{energy harvesting}
\newacronym{bs}{BS}{base station}
\newacronym{swipt}{SWIPT}{simultaneous wireless information and power transfer}
\newacronym{cots}{COTS}{commercial off-the-shelf}
\newacronym{dso}{DSO}{Distribution System Operator}
\newacronym{roi}{RoI}{return of investment}

\makeglossaries

\begin{document}
\pagenumbering{gobble}


\title{6G for Bridging the Digital Divide: \\Wireless Connectivity to Remote Areas}

\author{{{Abdelaali Chaoub},~\IEEEmembership{Senior Member, IEEE},
{Marco Giordani},~\IEEEmembership{Member, IEEE},
{Brejesh Lall},~\IEEEmembership{Member, IEEE},\\
{Vimal Bhatia},~\IEEEmembership{Senion Member, IEEE},
{Adrian Kliks},~\IEEEmembership{Senior Member, IEEE},
{Luciano Mendes},\\
{Khaled Rabie},~\IEEEmembership{Senior Member, IEEE},
{Harri Saarnisaari},~\IEEEmembership{Senior Member, IEEE},
{Amit Singhal},~\IEEEmembership{Member, IEEE},\\
{Nan Zhang},~\IEEEmembership{Member, IEEE},
{Sudhir Dixit},~\IEEEmembership{Life Fellow, IEEE},
{Michele Zorzi},~\IEEEmembership{Fellow, IEEE}}

\thanks{
Abdelaali Chaoub is with the National Institute of Posts and Telecommunications (INPT), Morocco (email: chaoub.abdelaali@gmail.com).
Marco Giordani and Michele Zorzi are with the Department of Information Engineering, University of Padova, Padova, Italy (email: \{giordani, zorzi\}@dei.unipd.it).
Brejesh Lall is with the Indian Institute of Technology Delhi, India (email: brejesh@ee.iitd.ac.in).
Vimal Bhatia is with the Indian Institute of Technology Indore,  India (email: vbhatia@iiti.ac.in).
Adrian Kliks is with the  Poznan University of Technology’s Institute of Radiocommunications, Poland (email: adrian.kliks@put.poznan.pl).
Luciano Mendes is with the National Institute of Telecommunications (Inatel), Brazil (email: luciano@inatel.br).
Khaled Rabie is with the Manchester Metropolitan University, UK (email: k.rabie@mmu.ac.uk).
Harri Saarnisaari is with the University of Oulu, Finland (email: harri.saarnisaari@oulu.fi).
Amit Singhal is with the  Bennett University, India (email: singhalamit.iitd@gmail.com).
Nan Zhang is with the Department of Algorithms, ZTE Corporation (email: zhang.nan152@zte.com.cn).
Sudhir Dixit is with the Basic Internet Foundation and University of Oulu (email: sudhir.dixit@ieee.org).
}
}

\maketitle

\begin{abstract}
In telecommunications, network service accessibility as a requirement is closely related to equitably serving the population residing at locations that can most appropriately be described as remote. Remote connectivity, however, would have benefited from a more inclusive consideration in the existing generations of mobile communications. To remedy this, sustainability and its social impact are being positioned as key drivers of sixth generation's (6G) research and standardization activities. 
In particular, there has been a conscious attempt to understand the demands of remote wireless connectivity, which has led to a better understanding of the challenges that lie ahead. In this perspective, this article overviews the key challenges associated with constraints on network design and deployment to be addressed for providing broadband connectivity to rural areas, and proposes novel approaches and solutions for bridging the digital divide in those~regions.
\end{abstract}

		\begin{tikzpicture}[remember picture,overlay]
\node[anchor=north,yshift=-10pt] at (current page.north) {\parbox{\dimexpr\textwidth-\fboxsep-\fboxrule\relax}{
\centering\footnotesize This paper has been accepted for publication in IEEE Wireless Communications, \textcopyright 2021 IEEE.\\
Please cite it as: A. Chaoub, M. Giordani, B. Lall, V. Bhatia, A. Kliks, L. Mendes, K. Rabie, H. Saarnisaari, A. Singhal, N. Zhang, S. Dixit, M. Zorzi, \\ “6G for Bridging the Digital Divide: Wireless Connectivity to Remote Areas”, IEEE Wireless Communications, 2021.}};
\end{tikzpicture}

\section{Introduction} 
\label{sec:introduction}

In 2018, 55\% of the global population  lived in urban areas. Further, 67\% of the total world's population had a mobile subscription, but only 3.9 billion people were using Internet, leaving 3.7 billion unconnected, with many of those living in remote or rural areas~\cite{ouluwhitepaper2020}. 
People  in these regions are not part of the information era and this digital segregation imposes several restrictions to their daily lives and prospects. Children growing up without access to the latest communication technologies and online learning tools are unlikely to be competitive in the job and commercial markets. Unreliable Internet connections also make it difficult for people living in remote areas to benefit from online commerce and engage in the digital world, thereby compounding already existing social and economic~inequalities. 

However, rural areas are now becoming more and more attractive  as the new coronavirus (COVID-19) pandemic has shown, since it has reshaped our living preferences and pushed many people to work remotely from wherever makes them most comfortable~\cite{phillipson2020covid}. Such agglomerations where people live and work are referred to as ``oases" in this paper. Wireless connectivity in  rural areas is  expected to have a significant economic impact too. The use of technology in farms or fisheries will open new opportunities for local communities (e.g., indigenous population in the US). Technology will also provide better education, increased digital social engagement, and efficient health systems to those living in the most remote zones (e.g., urban-rural divide in China and India).

Despite these premises, advances in the communication standards towards provisioning of wireless broadband connectivity to remote regions have  been, so far, relegated to the very bottom, if not entirely ignored.
In these regards, despite being in its initial stages, the \gls{6g} of wireless networks is building upon the issues left open by the previous generations~\cite{giordani2020toward}, and will be developed by taking into account the peculiarities of the remote and rural sector, with the objective of providing connectivity for all and reach digital~inclusion~\cite{yaacoub2020key}. Specifically, the research community should ensure that this critical market segment is not overlooked in favor of  the more appealing research areas such as \gls{ai}, \gls{ml}, Terahertz communications, 3D \gls{ar} and \gls{vr}, and~haptics.

\begin{figure*}[t!]
\centering
\includegraphics[width=.99\textwidth]{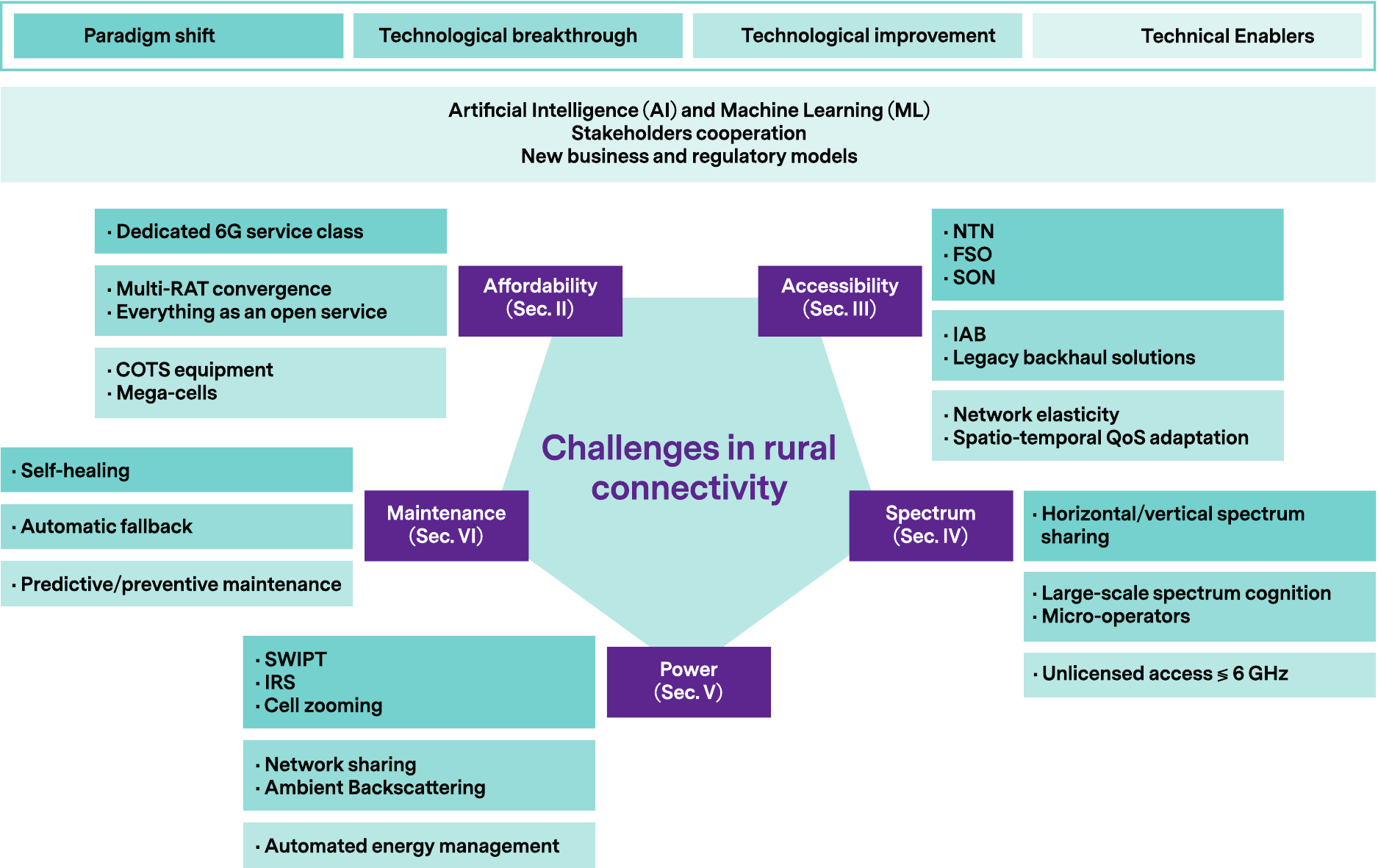}
    \caption{A summary of the challenges for providing wireless connectivity to remote areas with proposed solutions.}
    \label{fig:tab1}
\end{figure*}

Boosting remote connectivity can start by addressing spectrum availability issues. Licensed spectrum in sub-1 GHz bands, in fact, is a costly resource, and may require new frequency reuse strategy in remote regions because of their unique requirements. Judicious utilization of unlicensed bands may help reduce the overall cost, while advanced spectrum sharing models and enhanced co-existence schemes are two other powerful solutions to improve network coverage in these areas. Innovative business and regulatory models may be suitable to encourage new players, such as community-based micro-operators, to build and operate the local networks. Flexible and pluralistic spectrum licensing could be the way forward to boost the remote market.

Another issue is that remote areas may not have ample connectivity to the power sources (e.g., sub-Saharan Africa). Hence, it is imperative that 6G solutions be designed as self-reliant in terms of power/energy requirements, so as to improve \gls{roi} for remote areas, and/or with the ability to scavenge from the surrounding, possibly scarce, resources. Governments can assuage this situation to an extent by making it attractive for the profit-wary service providers to deploy solutions in remote areas. Revised government policies and appropriate business models should be jointly explored as they have direct implications on the technology requirements. Environmentally-friendly thinking should also be included throughout the chain from mining to manufacturing and recycling. Moreover, the abundant renewable sources need to be integrated into power systems at all scales for sustainable energy provisioning. 

Remote maintenance of network infrastructures is also very important since it might be difficult to access remote areas due to harsh weather and terrain, or lack of transport connectivity (e.g., in Arctic regions). Suitable specifications for fault tolerance and fallback mechanisms need therefore to be incorporated. 


%
Based on the above introduction, the fundamental challenges in remote areas are low return on investment, inaccessibility that hinders deployment and regular maintenance of network infrastructures, and lack of favorable spectrum and critical infrastructure such as backhaul and power grid, respectively. 
Along these lines, the contributions of this article are three-fold. First, we introduce trending technologies that future 6G networks should embrace from the early design and specification stages, to build a global connectivity for all. Second, we highlight the challenges that hinder progress in the development of solutions for remote areas. Notably, we grouped them into five categories: affordability (Sec.~\ref{sec:Affordable Service Provisioning in Remote Segments}), accessibility (Sec.~\ref{sec:improving_service_accessibility_in_under_served_areas_}), spectrum (Sec.~\ref{sec:towards_a_flexible_use_of_spectrum_in_remote_areas}), power (Sec.~\ref{sec:powering_in_a_green_and_efficient_way_} ), and maintenance (Sec.~\ref{sec:intelligent_and_affordable_maintenance_}). Third, we map those challenges into promising technologies for bringing global connectivity to beyond 5G networks. This includes the evolution of existing 5G innovations, as well as some new research ideas that will emerge within the next decade. A summary of these rural-specific challenges and possible solutions is shown in Fig.~\ref{fig:tab1}, while the main insights are provided in the following sections. We deliberately skip a detailed literature survey, because a clear and comprehensive review is provided in~\cite{yaacoub2020key}.

\begin{figure*}[t!]
\centering
\includegraphics[width=.99\textwidth]{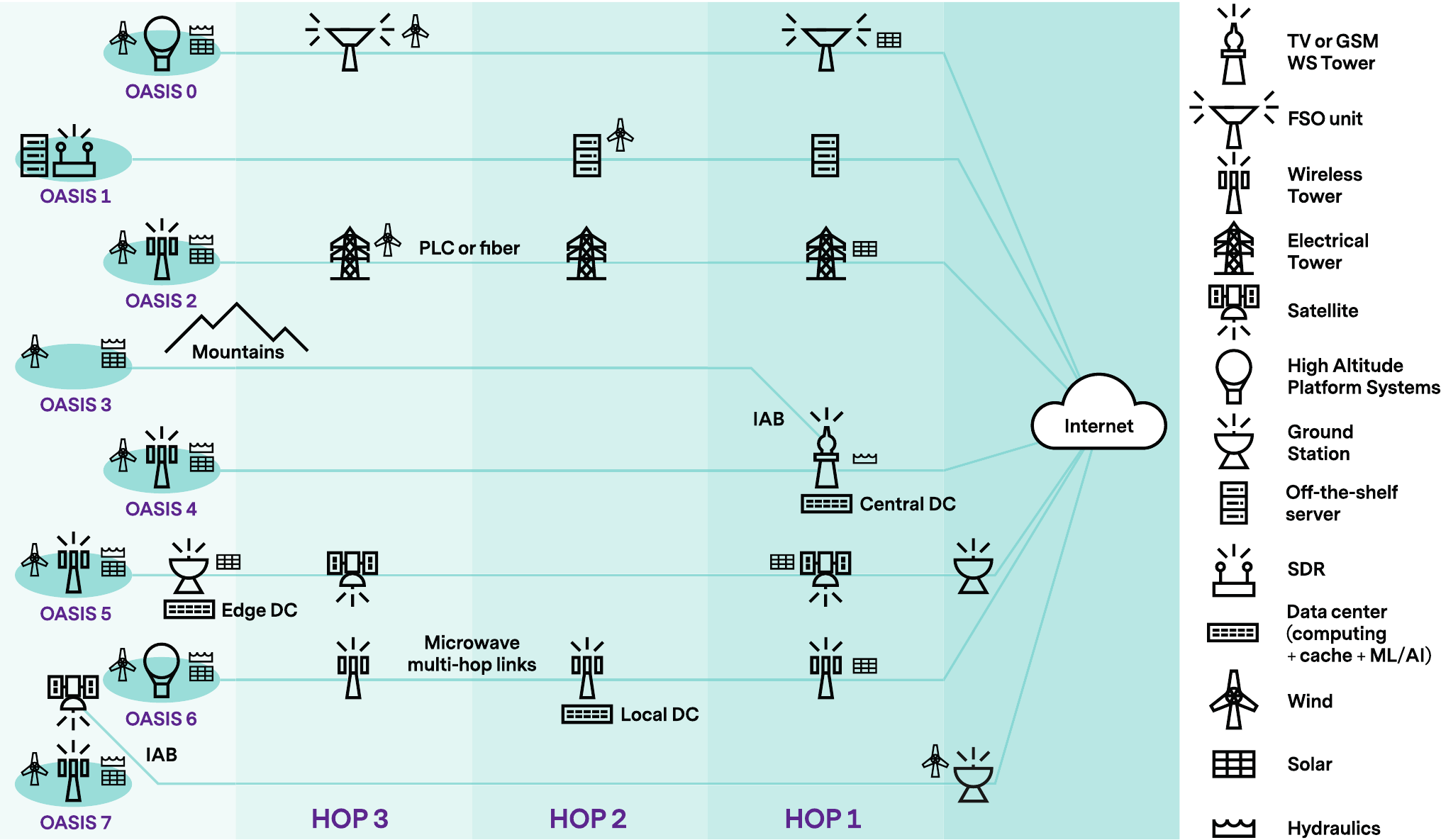}
    \caption{Backhaul connectivity in rural areas.}
    \label{fig:backhaul}
\end{figure*}
\section{Affordable Service Provisioning \\ in Remote Regions} 
\label{sec:Affordable Service Provisioning in Remote Segments}

Among the biggest impediments to connecting the unconnected part of the world are the high costs involved and the prevailing low income of the target population. Fortunately, there are many affordable emerging alternatives in 6G which may bring new possibilities, as listed in this section.

\subsection{Dedicated Remote-Centred Connectivity Layer} 
Besides 5G's typical service pillars, 6G should introduce a fourth service grade with basic connectivity requirements~\cite{yaacoub2020key}. However, this remote mode cannot be just a plain version of the urban 6G, since it has to be tailored to the specificities of the remote locations. Some demands relevant to remote connectivity scenarios like coverage and cost effectiveness need to be expanded, whereas the new service class needs more relaxed constraints in terms of some conventional 5G performance metrics like throughput and latency. This novel service class should have its dedicated slice, and be endowed with specific and moderate levels of edge and caching capabilities: the involved data can then be processed on edge, local or central data centers for better scalability, as illustrated in Fig.~\ref{fig:backhaul}. Accordingly, such connectivity services can be charged at reduced prices.

\subsection{Multiple \Glspl{rat} Interworking} 
\label{ssec:rat}
Local access in remote areas can be designed to aggregate multiple and heterogeneous \glspl{rat}. 
At the same time, digitalization in remote areas calls for large coverage solutions (e.g., TV or GSM \glspl{ws}), referred to as mega-cells, to increase the number of users within a base station and reduce the network deployment and management costs, albeit at some performance trade offs and with revised power limits.

\begin{figure*}[t!]
\centering
\includegraphics[width=.9\textwidth]{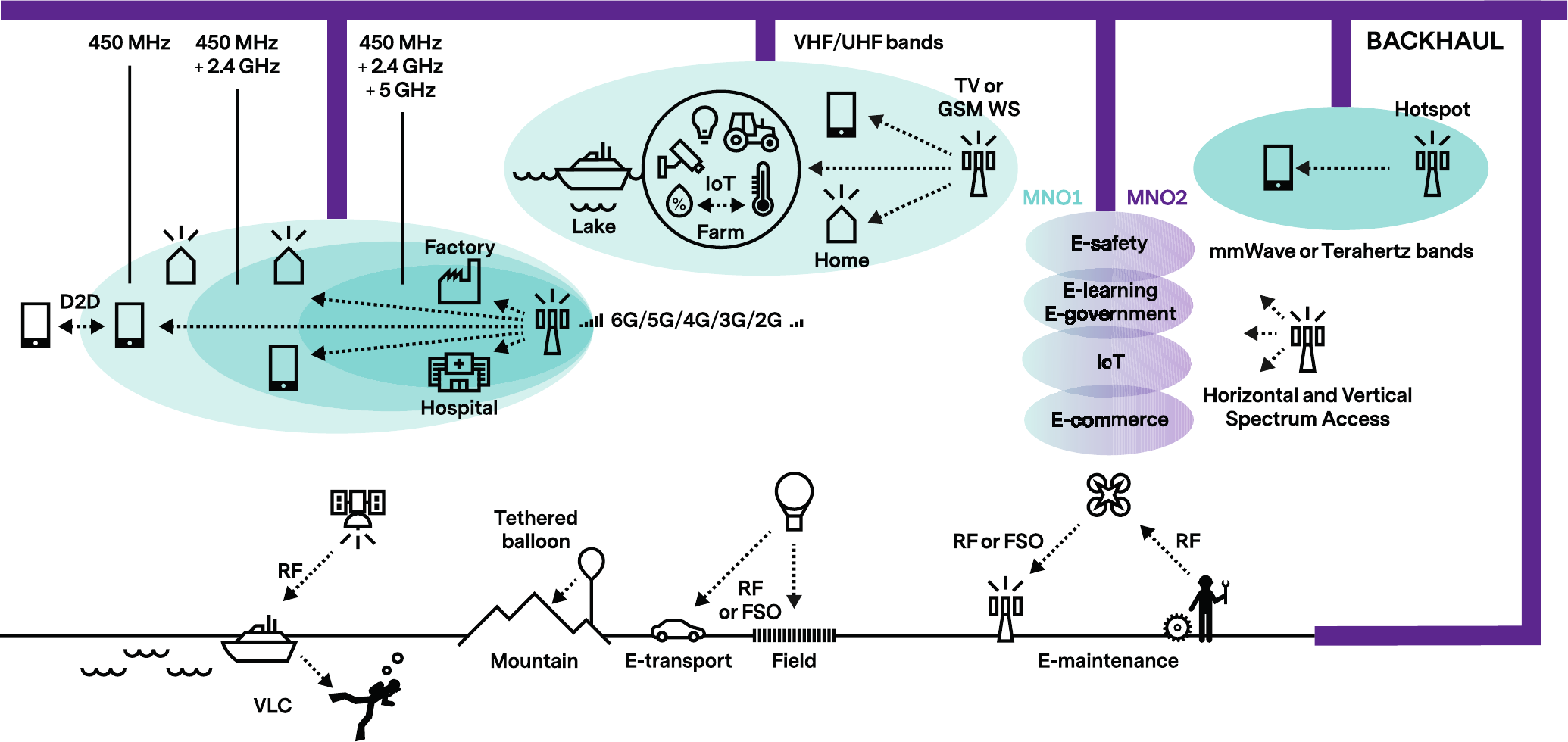}
    \vspace{-0.3cm}\caption{Spectrum usage for remote connectivity use cases.}
    \label{fig:remote}
\end{figure*}

Compared to 5G-like multi-RAT implementations, the availability of Terahertz and optical spectrum solutions in 6G will increase the heterogeneity and scale of the network, and open new opportunities for improving network performance~\cite{giordani2020toward}. In particular, \glspl{vlc} can boost the throughput indoor (i.e., where people spend 90\% of their time), fronthaul, and underwater environments (see Fig.~\ref{fig:remote}) while serving the intuitive goal of illumination, making it a cost-efficient technology.

\subsection{Low-Cost Networking and End-User Devices} One way to reduce cost is the exploitation of legacy infrastructure. TV stations can be shared with \glspl{mno} to provide both tower and electricity. The latest developments in wireless communications can be applied in outdoor \gls{plc} to provide high data rate connectivity over the high- and medium-voltage power lines, increasing the capability of the backhaul networks in remote areas. Existing base stations and the already-installed fibers alongside roads or embedded inside electrical cables can also serve as a backhaul solution for  connectivity in rural regions. 
End-user devices and modems should also be affordable and usable everywhere, i.e., when people move or travel to different places under harsh conditions. Therefore, the possibility to use \gls{cots} equipment at both the user's and network's sides is important, and integration with appropriate software stacks is welcome to reduce capital and operational expenditures (capex and opex).
For example, in December 2020, Facebook's SuperCell platform, integrating high-altitude towers with \gls{cots} radios, was deployed to improve the \gls{roi} for rural area network coverage. While one SuperCell can reduce costs of ownership up to 33\% compared to many equivalent macrocells, on the downside this prototype is power-hungry and presents a single point of failure for a large region.

\subsection{Adoption of Open/Virtualized/Cloud-Native Solutions} The remote infrastructure is likely to be deployed by small \glspl{isp} and the cost of specialized hardware equipment is an issue to be overcome. Open source approaches allow \glspl{mno} to choose common hardware from any vendor and implement the \gls{ran} and core functionalities using \gls{sdr} and \gls{sdn} frameworks. Moreover, virtualized and cloudified network functions may reduce infrastructure, maintenance and upgrade costs~\cite{gavrilovska2020cloud}. These solutions are especially interesting for new players building the remote network from scratch, to foster the inter-operability and cost-effectiveness of hardware and software. However, current 5G studies have not yet clarified whether virtualized equipment can be deployed at low cost, a critical requirement for rural network operators with limited access to economic resources.

\section{Improving Service Accessibility\\ in Remote Areas } 
\label{sec:improving_service_accessibility_in_under_served_areas_}

In order to provide long-lived broadband connectivity, a minimum service quality must be continuously guaranteed. In this perspective, this section reviews potential solutions to promote resilient service accessibility in rural~areas.

\subsection{Multi-Hop Network Elasticity} The access network has, over generations, become multi-hop to provide flexibility in the architecture design, despite some increase in complexity. Given the typical geographic, topographic, and demographic constraints of present scenarios, performance levels (e.g., coverage, latency, and bandwidth) of individual hops can be made adaptive. The idea is to extend performance elasticity beyond just the air interface to include other hops in the \gls{ran}. The same approach can be brought to backhaul connections (see Fig.~\ref{fig:backhaul}). 
Similarly, rural cell boundaries experiencing poor coverage can reap the elasticity benefits through the use of \gls{d2d} communications as depicted in Fig.~\ref{fig:remote}.

Network protocols should be extended to include spatial- (e.g., location-based) besides temporal-quality adaptation to handle variations in channel quality over time.

\subsection{Non-Terrestrial Networks} Network densification in rural areas is complicated by the heterogeneous terrain that may be encountered when installing fibers between cellular stations. To solve this issue, unlike current two-dimensional 5G networks, 6G envisions the deployment of \glspl{ntn} in 3D through air/spaceborne platforms like \glspl{uav}, \glspl{haps}, and satellites (e.g., \gls{leo} constellations for last-mile connectivity)~\cite{giordani2020non}. \glspl{ntn} can offer a robust standalone networking solution to preserve connectivity in the absence of other already-deployed network infrastructures, or when terrestrial towers are out of service, e.g., after natural disasters~\cite{cheng2020comprehensive}. 
Potential beneficiaries of this trend are shown in Fig.~\ref{fig:remote}, including inter-regional transport, farmlands, ships, mountainous areas, and remote maintenance facilities. The evolution towards \glspl{ntn} will be favored by architectural advancements in the aerial/space industry (e.g., through solid-state lithium batteries and Gallium Nitride technologies), new spectrum developments (e.g., by transitioning to \gls{mmwave}, Terahertz and optical bands), and novel antenna designs (e.g., through reconfigurable phased/inflatable/fractal antennas realized with metasurface material). Despite these premises, however, there are still various challenges to be addressed, including those related to latency, coverage and energy constraints. For example, emerging \gls{leo} satellite initiatives like Starlink by SpaceX can provide global coverage and low-latency services once operational. Nevertheless, how to deal with end-of-life disposal of satellites and constellation management is yet to be addressed.

\subsection{Wireless Backhaul} 
Service accessibility in rural areas  involves  prohibitive  deployment  expenditures  for  network  operators  and  requires  high-capacity  backhaul  connections  for  several  different  use  cases. Fig.~\ref{fig:backhaul} provides a comprehensive overview of potential backhaul solutions envisioned in this paper to promote remote connectivity. On one side, laying more fiber links substantially boosts broadband access in those areas, but at the expense of increased costs. \gls{plc} connections, on the other side, provide ease of reach at lower costs making use of  ubiquitous wired infrastructures as a physical medium for data transmission, but some inherent challenges related to harsh channel conditions are still to be overcome. 
Fig.~\ref{fig:backhaul} illustrates also how
emerging long-range wireless technologies, such as TV and GSM \gls{ws} systems, are capable of delivering the intended service over longer distances with less power while penetrating through difficult terrain like mountains and lakes. \Gls{ntn}, as foreseen in 6G, could also provide a remote-ready,  low-cost (yet robust), and long-range backhaul alternative to complement conventional technologies in remote areas.

Another recent trend is building cost-effective backhaul using software-defined technology to connect the unconnected communities (e.g., Oasis 1 in Fig.~\ref{fig:backhaul}). Recently, the  research  community  has  also investigated \gls{iab} as a solution to replace  fiber-like infrastructures  with  self-configuring  relays operating through wireless backhaul~\cite{polese2020integrated}. 
For example, the TV \gls{ws} tower in Fig.~\ref{fig:backhaul} may use the TV spectrum holes to provide both access to Oasis 3 and connection to the backhaul link for Oasis 4. IAB has lower complexity compared to fiber-like networks and facilitates site installation in rural areas where cable buildout is difficult and costly. 
As a case study, in 2020 Huawei's RuralStar Pro prototype, an \gls{iab}-based solar-powered lightweight base station, was deployed in a mountainous village of China. The site installation was 80\% faster than a traditional network, and improved the LTE coverage by more than 80\% with only 100 W power consumption.
While, in 5G, wireless backhaul is typically realized at \glspl{mmwave}, the huge capacity of 6G-specific technologies, e.g., Terahertz, could be exploited to better multiplex the access and backhaul data within the same bands, thereby removing the need for additional hardware and/or spectrum licence costs.

Nowadays, \gls{fso} links are also being considered as a powerful alternative to increase network footprint in isolated areas with challenging terrains. However, FSO units are very sensitive to optical misalignment. 
For instance, the HOP1 FSO unit depicted in Fig.~\ref{fig:backhaul} should be permanently and perfectly aligned with the FSO unit installed in the HOP3 location. 
In-depth research on spherical receivers and  beam  scanning  is  hence needed  to  improve  the  capability  of intercepting laser lights emanating from multiple angles.

\subsection{Physical-Layer Solutions for Front/Mid/Backhaul} Even though wireless backhauling can reduce deployment costs, service accessibility in rural regions still requires a minimum number of fiber infrastructures to be already deployed. Fiber capacity can hence be increased if existing wavelength division multiplexing networks are migrated to \glspl{eon} by technology upgrades at nodes; the outdated technology of urban regions may then be reused to establish connectivity in under-served rural regions without significant~investment. 

Besides backhaul, midhaul and fronthaul should also be improved by \gls{ai}/\gls{ml}-based solutions providing cognitive capabilities for prudent use of available licensed and unlicensed spectrum~\cite{li2020collective}. This is especially useful in remote areas where the sparse distribution of users may result in spectrum holes. The unlicensed spectrum, in particular, can promote better RoI for service delivery and improve network elasticity, as will be discussed in Sec.~\ref{ssec:unlicensed}.
 New possibilities including evolved multiple access schemes and waveforms, like \gls{noma}, should be investigated; this technology is particularly interesting for \gls{iot} services where some sensors are close to and some far away from a base station~\cite{hu2019application}. \gls{ai}/\gls{ml} can be exploited to control physical and link layers for smooth and context-aware \glspl{mcs} transitions, even though this approach would need to be lightweight to reduce cost and maintenance, and optimized for the intended market segment.

\subsection{\Glspl{son}} To explicitly address the problem of network outages (e.g., due to backhaul failure),  which are very common in remote locations, 6G should transition towards \glspl{son} implementing network slicing, dynamic spectrum management, edge computing, and zero-touch automation functionalities. This approach  provides extra degrees of freedom for combating service interruptions, and improves network robustness.
In this context, \gls{ai}/\gls{ml} can help both the radio access and backhaul networks to self-organize and self-configure themselves, e.g., to discover each other, coordinate, and manage the signaling and data traffic~\cite{li2020collective}.

\section{Towards a Flexible Use of Spectrum \\in Remote Areas} 
\label{sec:towards_a_flexible_use_of_spectrum_in_remote_areas}

We now present some promising solutions to address spectrum availability issues, which currently pose a serious impediment to broadband connectivity in  remote areas.  

\subsection{Cognitive Radio Networks Using Unlicensed Bands} 
\label{ssec:unlicensed}
One  of  the  major barriers  for  network  deployment  in  rural  areas  is  spectrum licensing,  since  participation  in  spectrum  auction  is  hindered to small  \glspl{isp}. Indeed, 6G may evolve towards fully autonomous standalone networks that uniquely operate in unlicensed bands, for example where  giant  operators are  not  interested  in  providing  their  service. In these regards, the unlicensed access variants of 5G mainly operate in the (already crowded) low/mid bands, and do not integrate 6G-specific bands, thus complicating network operations. The massive connectivity demands of rural vertical industries (e.g., farming, mining, etc.) will further exacerbate this issue.

In this perspective, 6G should consider several other frequencies for unlicensed operations, e.g., in the 90--100 GHz and Terahertz bands.
 Moreover, existing synchronization and control mechanisms like \gls{lbt} should mature towards more distributed intelligence-empowered algorithms to avoid misleading spectrum occupancy, and better orchestrate network operations. Further, endowing 6G mobile radios with large-scale cognition capabilities, along with autonomous and intelligent tuning of radio parameters, will allow  ISPs to instantaneously obtain an accurate spectrum occupancy map, and be informed about vacant spectrum opportunities in a given area. Such an approach will also provide protection  against unauthorized  transmissions  and unpredictable propagation conditions to new~entrants.

\subsection{Spectrum Co-Existence} Sub-6 GHz frequencies remain critical for remote connectivity thanks to their favorable propagation properties and wide reach, confirmed by many successful rural/remote broadband initiatives worldwide (e.g., UHF/VHF TV \glspl{ws} in Africa,  CDMA 450 MHz in Ecuador, WiFi 2.4 and 5 GHz in Asia).
In these crowded bands, spectrum re-farming and inter/intra-operator spectrum sharing can increase spectrum availability~\cite{ahmad20205g}. 
Nevertheless, coverage gaps and low throughput in the legacy bands call for expanding the operational bandwidth of 5G systems over different spectra, combining frequencies above and below 6 GHz besides the optical spectrum. Using advanced carrier aggregation techniques in 6G systems, the resource scheduling unit can choose the optimal frequency combination(s) according to service requirements and spectrum characteristics. The proposed model offers a scalable bandwidth that maintains service continuity in case of connectivity loss in those spectrum bands that are more sensitive to surrounding effects. 
Likewise, multi connectivity (as discussed in Sec.~\ref{ssec:rat}) provides diversity, improved system resilience, and situation awareness by establishing multiple links from separate sources to one destination. This aggregation can be achieved at various protocol and/or architecture levels ranging from the radio link up to the core network, allowing effortless deployment of elastic networks in areas difficult to access.


\subsection{Regional Licenses and Micro-Operators} Deployment of terrestrial networks for remote areas is challenging due to harsh terrain and lack of infrastructure and personnel. Network operators would then rather roam their services from telecommunication providers already operating in those areas than building their own infrastructure. However, such an approach may entail the need for advanced horizontal (between operators of the same priorities) and vertical (when stakeholders of various priorities coexist) spectrum/infrastructure sharing frameworks. Solutions like license shared access (LSA, in Europe) and spectrum access system (SAS, in the US) are mature examples of such an approach with two tiers and three tiers of users, respectively. This can evolve to include $n$ tiers of users belonging to $m$ different \glspl{mno}. An example of a four-tiered access is provided in Fig.~\ref{fig:remote}. From the top, we find the E-safety services with the highest priority, a tier-2 layer devoted to E-learning sessions and E-government transactions, a  middle-priority tier-3 layer for \gls{iot} use cases that generate sporadic traffic, and a final lower-priority tier-4 layer that uses the remainder of the available spectrum (e.g., for E-commerce services). Such solutions, however, need to be supported by innovative business and regulatory models to motivate new market entrants (e.g., micro-operators) to offer competitive and affordable services in remote zones~\cite{ahokangas2019business}.

\begin{figure}[t!]
\centering
\includegraphics[width=.9\columnwidth]{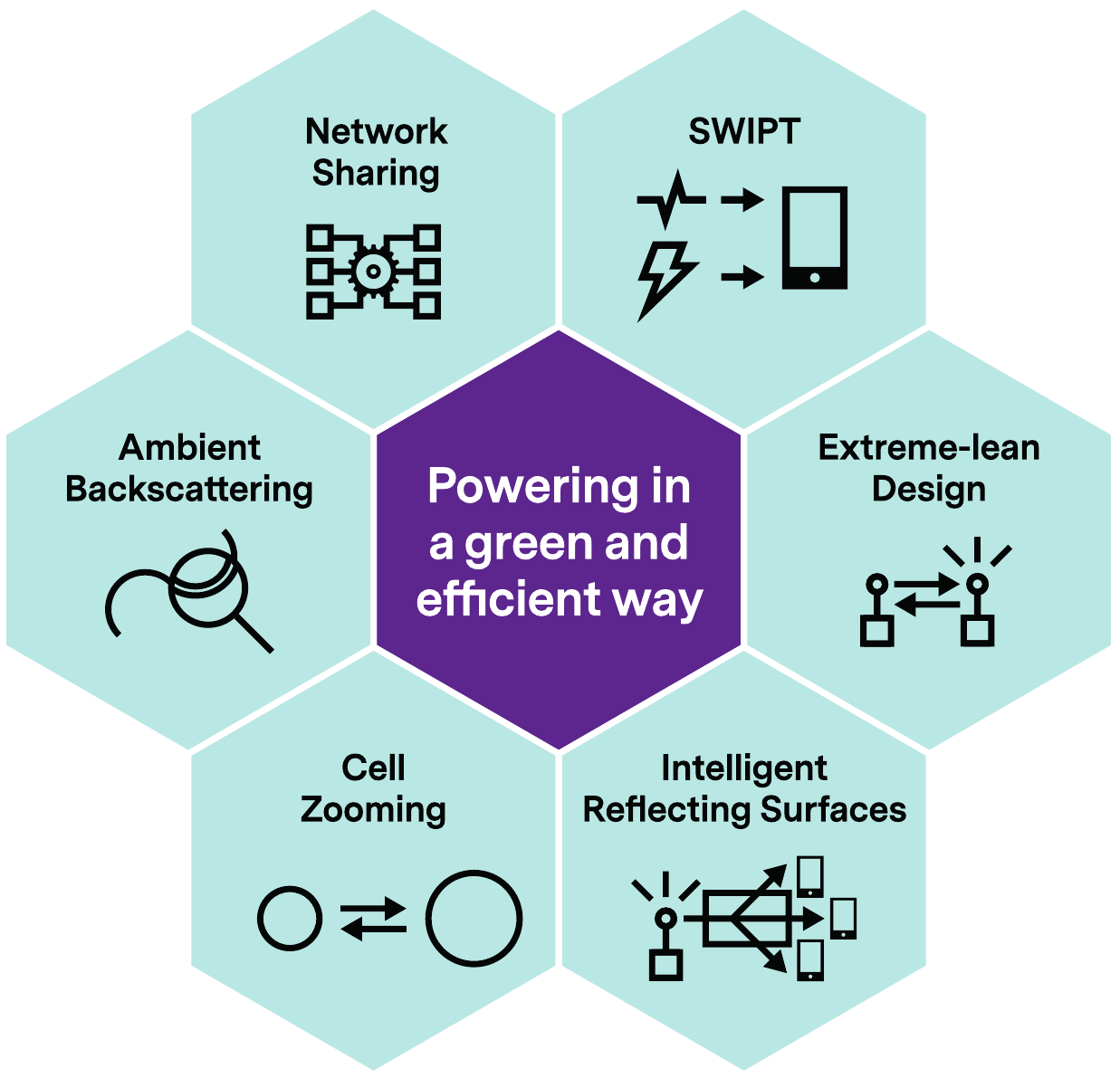}
    \caption{Key enablers for efficient and green powering of rural areas.}
    \label{fig:powering}
\end{figure}

\section{Powering in a Green and Efficient Way} 
\label{sec:powering_in_a_green_and_efficient_way_}

Power supply is among the highest expenses of \glspl{mno} and a major bottleneck for ensuring robust connectivity in remote areas. 
\glspl{mno}' reliable powering can be improved following (a combination of) these solutions, as summarized in Fig.~\ref{fig:powering}.

\subsection{Efficient and Optimal Energy Usage} Energy efficiency  while guaranteeing affordable yet robust connectivity for residents remains a key requirement for 6G remote solutions. Endowing rural networks with energy metering capabilities at the component level is a driver to support autonomous energy management systems, while software-enabled renewable-powered units will help attain significant power.

Techniques like cell zooming relying on power control and adaptive coverage, that were deliberately left out of early 5G standards development, can also enhance RoI for front/mid/backhaul layouts, e.g., by adaptively zooming out/switching off base stations deployed where the population density is particularly low. Multi-connectivity schemes can also harness cell zooming at cell edge and shadowed locations for expanded coverage and capacity.

\subsection{Network Sharing} Telecom operators and stakeholders (e.g., regional content providers, manufacturers, governmental authorities, etc.) should coordinate a joint network development process right from the early 6G standardization stages. Network design shall foresee different levels of sharing, ranging from infrastructure/network to multi-tenant resource sharing. In particular, the different players should cooperate to avoid deploying separate power plants and hardware for those services offered in the same rural area, thus saving precious (already limited) economic resources for other types of expenses. For network sharing agreements to be a success, 6G needs to guarantee full competition, independent control, and secure~services.

\subsection{Technological Breakthroughs} The \gls{swipt} paradigm enables energy-constrained nodes to scavenge energy from information signals. Incorporated in 6G, this technology can foster cheap  manufacturing of battery-free sensors for rural services with intermittent traffic (e.g., agriculture). 
On one side, time-switching-based \gls{swipt}, which is technically less complex than power switching, can potentially provide sufficient throughput to sparsely populated regions. Information bits can be conveyed through backscattering ambient signals for short-range scenarios.
At the same time, \gls{swipt} platforms can be installed in rural locations (e.g., on flying platforms) to charge low-power network elements starved of energy. The low levels of harvested energy, however, are still impeding the integration of advanced power-hungry functions in wireless-powered networks, thus calling for further research~\cite{swipt8214104}.

Another revolutionary element of the 6G ecosystem promoting energy-efficient wireless operations is the use of \glspl{irs}, equipped with a large number of passive elements smartly coordinated to reflect any incident signal to its intended destination without necessitating \gls{rf} chains. On one side, passive beamforming may result in significant power loss over long distance, preventing standalone coverage in remote areas. In turn, \glspl{irs}, combined with existing infrastructures, can alter the electromagnetic environment experienced by the users, and favorably relay the traffic in underconnected regions, e.g., at the network edge, thus achieving higher overall network performance with substantial energy savings~\cite{wu2019towards}.

\section{Intelligent and Affordable Maintenance} 
\label{sec:intelligent_and_affordable_maintenance_}

\begin{figure}[t!]
\centering
\includegraphics[width=.99\columnwidth]{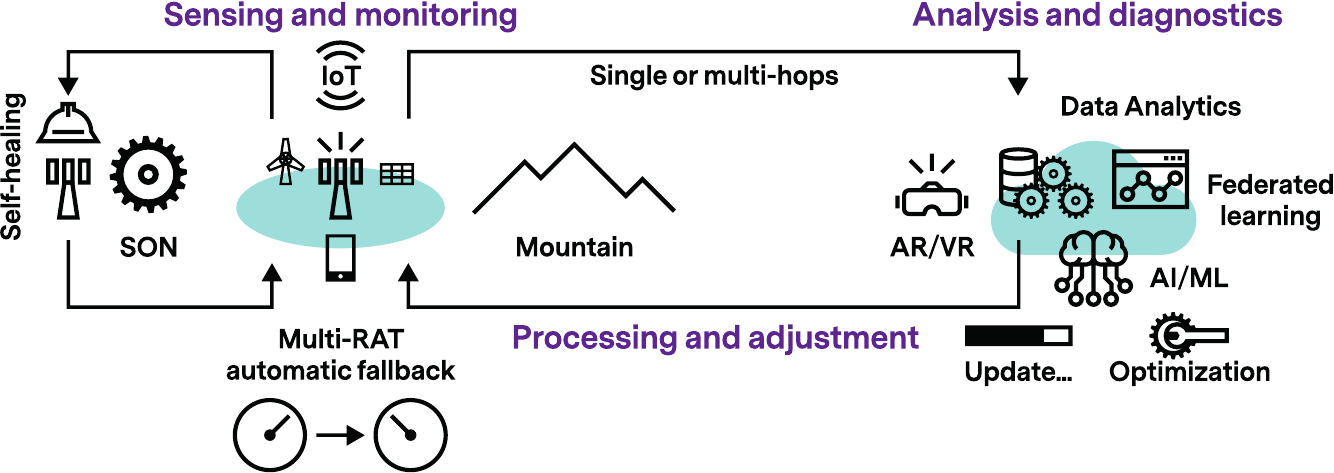}
    \caption{A workflow for cost-efficient maintenance operations in remote areas.}
    \label{fig:maintenance}
\end{figure}

Hard-to-reach areas increase the complexity of maintenance in commercial 4G/5G networks, which is further exacerbated by the lack of qualified personnel residing is those areas. In this section we present innovative ideas to enable intelligent and cost-effective maintenance in 6G networks deployed in rural regions, as depicted in Fig.~\ref{fig:maintenance}.

\subsection{Network Intelligence and Automation} In  the  next  decade,  6G  is expected to leverage the power of collective intelligence by gathering huge volumes of data to  prevent network malfunction and self-optimize~\cite{li2020collective}. This will minimize the need for on-site intervention in remote  facilities and truly operate the network in an autonomous fashion. Nevertheless, such services will increase the need for centralized data processing in rural locations, which may become a bottleneck in large-scale scenarios.
A promising approach is to design distributed learning strategies capable of offloading complex computation tasks into decentralized environments with abundant energy supply. Federated learning may further simplify network automation by processing learning models (e.g., on cheap hardware and/or flying \gls{sdr} platforms, as recommended for rural connectivity) instead of training data~\cite{li2020collective}.

\subsection{Proactive Outage Detection} \Gls{ar} and \gls{vr} environments can remotely simulate real scenarios where a radio component runs under typical remote/rural conditions. Outcomes of these simulations can be combined with both external historical information of similar components, and internal contextual data gathered using pervasive intelligence, to predict node failures. This will help 6G networks build solid and automatic backup plans (e.g., system updates or operational parameter optimization) for hard-to-reach areas where on-site repair activities could be impractical~\cite{verde2020advanced}.
\smallskip

\subsection{Self-Healing Capabilities} Intermittent power availability and battery-constrained equipment in rural locations can lead to connectivity failures. 
Reflecting panels, e.g., \glspl{irs}, mounted on aerial platforms are able to achieve panoramic signal reflection from the sky and temporarily replace the front/backhaul for remote nodes. Further, they can direct \gls{rf} energy towards low-power devices like farming \gls{iot} sensors for energy harvesting. Automatic fallback can also be scheduled to downgrade the connectivity to another technology under poor network conditions, e.g., by implementing appropriate multi-connectivity schemes as described in Sec.~\ref{ssec:rat}. All these capabilities will benefit from the strong support of 6G collective intelligence~\cite{li2020collective}.

\begin{table*}[t!]
\centering
\footnotesize
\setlength{\belowcaptionskip}{0.33cm}
\caption{Open research questions for catering to remote regions in 6G.}
\label{tab:challenges}
\renewcommand{\arraystretch}{1.5}
\begin{tabular}{|m{2cm}|m{5cm}|m{9.5cm}|}
\hline
\textbf{Challenge}  & \textbf{Potential Solutions} & \textbf{Future Research Directions} \\
\hline
\multirow{4}{*}{\begin{tabular}[c]{@{}l@{}}Affordability \\ (Sec.~\ref{sec:Affordable Service Provisioning in Remote Segments}) \end{tabular}} & Dedicated 6G service class & Define corresponding requirements and relevant \glspl{kpi}\\ 
& Multi-\gls{rat} convergence & Seamless resource orchestration over heterogeneous \gls{rf} and optical bands\\ 
& \gls{cots} equipment & Obsolescence concerns and need for anticipated replacement plans\\ 
& Everything as an open service & Inter-operability and cost-effectiveness concerns\\ 
& Mega-cells & New guidelines for safety and power limits \\ 
\hline
\multirow{7}{*}{\begin{tabular}[c]{@{}l@{}}Accessibility \\ (Sec.~\ref{sec:improving_service_accessibility_in_under_served_areas_}) \end{tabular}} & Network elasticity & Flexible and automated resource orchestration\\ 
& Legacy backhaul solutions & Integration with emerging backhaul solutions (e.g. \gls{iab}, \gls{ntn}, \gls{fso}) to reduce costs\\ 
& Spatio-temporal QoS adaptation & Intelligent tuning of encoding parameters to reduce complexity\\ 
& \gls{iab} & Balanced resource multiplexing between access and backhaul\\ 
& \gls{ntn} & Complicated placement in 3D space and limited energy concerns\\ 
& \gls{fso} & Sensitivity to optical misalignment\\ 
& \gls{son} & Evolving towards proactive self-coordinated 6G \glspl{son} using collective intelligence\\ 
\hline

\multirow{5}{*}{\begin{tabular}[c]{@{}l@{}}Spectrum \\ (Sec.~\ref{sec:towards_a_flexible_use_of_spectrum_in_remote_areas}) \end{tabular}} & Unlicensed access $\lessgtr$ 6 GHz &  Evolution towards collective intelligence-empowered control\\ 
& Large-scale spectrum cognition & Sharing limited local information/models instead of complex data\\
& Micro-operators & Innovative business and regulatory models encouraging new local entrants\\ 
&  Horizontal/vertical spectrum sharing & Need for rich environmental awareness for interference-free coexistence\\ 
\hline

\multirow{7}{*}{\begin{tabular}[c]{@{}l@{}}Power \\ (Sec.~\ref{sec:powering_in_a_green_and_efficient_way_}) \end{tabular}} & Network sharing & Full competition, independent control and secure services in multi-tenant networks\\ 
& Automated energy management & Component-centric energy metering capabilities with pervasive intelligence\\ 
& Ambient Backscattering & Short transmission range\\ 
& Cell zooming & Need to achieve a complex level of data-driven coordination\\ 
& \gls{swipt} & Low levels of harvested energy\\ 
& \gls{irs} & Channel estimation overhead and deployment complexity in 3D air-ground networks\\ 
\hline

\multirow{3}{*}{\begin{tabular}[c]{@{}l@{}}Maintenance \\ (Sec.~\ref{sec:intelligent_and_affordable_maintenance_}) \end{tabular}} & Predictive/preventive maintenance & Need for real-time computation offloading mechanisms\\ 
& Self-healing & Based on emerging technologies still maturing (e.g., \gls{irs}, \gls{uav}, energy harvesting)\\ 
& Automatic fallback & Preserving seamless and uninterrupted services during fallback\\
\hline
\end{tabular}
\end{table*}

\section{Conclusions} 
\label{sec:conclusions}

The problem of providing connectivity to rural areas will be a pillar of future 6G standardization activities. In this article we discuss the challenges and possible approaches to addressing the needs of the remote areas, as summarized in Table~\ref{tab:challenges}. 
It is argued that optimally integrating NTN and FSO technologies can provide low-cost broadband solutions in extremely harsh environments, and can be the next disruptive technology for 6G remote connectivity. Integration of outdated technologies should also be provisioned so that they may be used to service the remote areas. Such provisions should extend to integrate \gls{cots} solutions to fully benefit from cost advantage gains. Spectrum, regulatory, and standardization issues are also discussed to support remote area connectivity. 
We also expect forthcoming 6G to largely embrace multidisciplinary features, i.e., pervasive intelligence, stakeholders cooperation and flexible regulations, as key enablers for inclusive connectivity. 
These approaches, directly or indirectly, address the very fundamental challenge of \gls{roi} in deploying remote connectivity solutions. For this reason, the discussion in this article repeatedly alludes to capex/opex savings innovations which can be facilitated by appropriate technical, policy, and regulatory mechanisms in 6G.

While the support of existing cellular deployments may already provide preliminary coverage, resilience, and flexibility in rural regions, it is yet to be studied how standalone deployments could support remote connectivity where integration with existing communication technologies is impractical, e.g., on top of mountains, on the ocean, or in the air space.

\bibliographystyle{IEEEtran}

\begin{IEEEbiographynophoto}{Abdelaali Chaoub} [SM] is an Associate Professor, working at the National Institute of Posts and Telecommunications (INPT) at Rabat (Morocco) since 2015. His research interests are related to spectrum sharing for 5G/B5G networks, cognitive radio networks, smart grids, cooperative communications in wireless networks, and multimedia content delivery. He is a paper reviewer for several leading international journals and conferences. He has accumulated intersectoral skills through work experience both in academia and industry as a Senior VoIP solutions Consultant at Alcatel-Lucent (2007–2015).\end{IEEEbiographynophoto}%

\begin{IEEEbiographynophoto}{Marco Giordani} [M’20] received his Ph.D. in Information Engineering in 2020 from the University of Padova, Italy, where he is now a postdoctoral researcher and adjunct professor. He visited NYU and TOYOTA Infotechnology Center, Inc., USA. In 2018 he received the “Daniel E. Noble Fellowship Award” from the IEEE Vehicular Technology Society. His research focuses on protocol design for 5G/6G mmWave cellular and vehicular networks. \end{IEEEbiographynophoto}%

\begin{IEEEbiographynophoto}{Brejesh Lall} [M] received his Ph.D degree from Indian Institute of Technology Delhi in 1999. Earlier, he received his bachelors and masters degree in Electronics and Communications from Delhi College of Engineering in 1991 and 1992 respectively. He is currently a Professor in the department of Electrical Engineering at Indian Institute of Technology Delhi. Previously, he served in the Digital Signal Processing Group of Hughes Software Systems for 8 years. His research interests lie in the areas of signal processing and machine learning. He has extensively applied signal processing / machine learning techniques to applications in the broad areas of telecommunications and computer vision.\end{IEEEbiographynophoto}%

\begin{IEEEbiographynophoto}{Vimal Bhatia} [M’96, SM’12] is currently working as a Professor at Indian Institute of Technology Indore, India. He received his Ph.D. degree from the Institute for Digital Communications, The University of Edinburgh. During his Ph.D. studies he also received the IEE fellowship for collaborative research at Carleton University, Ottawa, Canada. He has authored/co-authored more than 270 peer-reviewed journals and conferences. His research focuses on optical communication networks, wireless communications, and machine learning. He is currently He is currently an Editor for IETE Technical Review, Frontiers in Communications and Networks, and IEEE Wireless Communications Letters.\end{IEEEbiographynophoto}%

\begin{IEEEbiographynophoto}{Adrian Kliks} [SM] is an assistant professor at Poznan University of Technology’s Institute of Radiocommunications, Poland, and he is a cofounder and board member of RIMEDO Labs company. His research interests include new waveforms for wireless systems applying either non-orthogonal or non-contiguous multicarrier schemes, cognitive radio, advanced spectrum management, deployment and resource management in small cells, and network virtualization.\end{IEEEbiographynophoto}%

\begin{IEEEbiographynophoto}{Luciano Mendes} received his Ph.D. on Electrical Engineering from the State University of Campinas, Brazil in 2007. Since 2001 he is a professor at the National Institute of Telecommunications (Inatel), Brazil, where he acts as Research Coordinator of the Radiocommunications Reference Center. His main research area is physical layer for future mobile networks.\end{IEEEbiographynophoto}%

\begin{IEEEbiographynophoto}{Khaled Rabie} [M’15, SM’20], a Fellow of the U.K. Higher Education Academy, received his Ph.D. degree from the University of Manchester, UK. He is currently Senior Lecturer at the Manchester Metropolitan University, UK. His primary research focuses on various aspects of the next-generation wireless communication systems. Khaled is an Editor of the IEEE Communications Letters, IEEE Internet of Things Magazine, and IEEE Access.\end{IEEEbiographynophoto}%

\begin{IEEEbiographynophoto}{Harri Saarnisaari} [SM] received his Ph.D degree from the University of Oulu in 2000, where he has been with Centre for Wireless Communications since 1994. He is currently a university researcher and his current research interest include remote area connectivity, especially in the Arctic areas.\end{IEEEbiographynophoto}%

\begin{IEEEbiographynophoto}{Amit Singhal} [M] received his PhD degree in electrical engineering from the Indian Institute of Technology Delhi in 2016. He is currently working as an Assistant Professor at Bennett University, Greater Noida, India. His research interests include next generation communication systems, Fourier decomposition method, image retrieval and molecular communications.\end{IEEEbiographynophoto}%

\begin{IEEEbiographynophoto}{Nan Zhang} [M] was born in Qingyang, China, in 1990. He received the bachelor degree in communication engineering and the Master degree in integrated circuit engineering from Tongji University, Shanghai, China, in July 2012 and March 2015, respectively. He is now a Senior Engineer at the Department of Algorithms, ZTE Corporation and works on the standardization of LTE and NR system. His current research interests are in the field of 5G channel modeling, MIMO, NOMA techniques, satellite/ATG communication and network architecture.\end{IEEEbiographynophoto}%

\begin{IEEEbiographynophoto}{Sudhir Dixit} [LF] is a Co-Founder of the Basic Internet Foundation in Oslo, Norway, and heads its US operations.  He is also a Docent at University of Oulu.  From 2015 to 2017 he was the CEO and Co-Founder of a start-up, Skydoot, Inc. in California. From 2009 to 2015, he was a Distinguished Chief Technologist and CTO of the Communications and Media Services for the Americas Region of Hewlett-Packard Enterprise Services in Palo Alto, CA, and the Director of Hewlett-Packard Labs India in Palo Alto and Bangalore. Before joining HP, he worked for BlackBerry, NSN, Nokia and Verizon. He has published 8 books, over 200 papers and holds 21 US patents. Dixit holds a Ph.D. from the University of Strathclyde, Glasgow, U.K. and an M.B.A. from the Florida Institute of Technology, Melbourne, Florida. \end{IEEEbiographynophoto}

\begin{IEEEbiographynophoto}{Michele Zorzi} [F] is with the Information Engineering Department of the University of Padova, focusing on wireless communications research. He was Editor-in-Chief of IEEE Wireless Communications from 2003 to 2005, IEEE Transactions on Communications from 2008 to 2011, and IEEE Transactions on Cognitive Communications and Networking from 2014 to 2018. He served ComSoc as a Member-at-Large of the Board of Governors from 2009 to 2011, as Director of Education and Training from 2014 to 2015, and as Director of Journals from 2020 to 2021.
\end{IEEEbiographynophoto}
\end{document}